\documentclass[fleqn,usenatbib]{mnras}

\usepackage{newtxtext,newtxmath}
\usepackage[T1]{fontenc}
\usepackage{ae,aecompl}
\usepackage{graphicx}	
\usepackage{amsmath}	
\usepackage{amssymb}	
\usepackage{gensymb}
\usepackage{float}

\def\beq{\begin{equation}}
\def\eeq{\end{equation}}

\def\Msun{$M_{\odot}$}

\title[A red UDG in a void]{Discovery of a red ultra-diffuse galaxy in a nearby void based on its globular cluster luminosity function}

\author[Rom\'an et al.]{Javier Rom\'an$^{1,2}$ \thanks{E-mail:jromanastro@gmail.com},  Michael A. Beasley,$^{1, 2}$, Tom\'as Ruiz-Lara$^{1,2}$ and David Valls-Gabaud$^{3}$\\
\newline \\
$^{1}$ Instituto de Astrof\'isica de Canarias, Calle V\'ia L\'actea s/n,
E-38205 La Laguna, Tenerife, Spain \\
$^{2}$ Departamento de Astrof\'isica, Universidad de La Laguna, E-38200 La
Laguna, Tenerife, Spain \\
$^{3}$LERMA, CNRS, PSL, Observatoire de Paris, 61 Avenue de l'Observatoire, 75014 Paris, France}

\date{Submitted, \today}

\begin{document}
\label{firstpage}
\maketitle

\begin{abstract}
Distance determinations of extremely-low-surface-brightness galaxies are expensive in terms of spectroscopic time. Because of this, their distances are often inferred by associating such galaxies with larger structures such as groups or clusters, leading to a systematic bias by selecting objects in high density environments. Here we report the discovery of a red ultra-diffuse galaxy (S82-DG-1: $r_{\mathrm{eff}}$~=~1.6~kpc; $<\mu _{g}>$~=~25.7~mag arcsec$^{-2}$; $g-i$~=~0.78~mag) located in a nearby cosmic void. We used multi-band luminosity functions of its globular clusters to obtain the distance to S82-DG-1, at 28.7$_{-3.6}^{+4.2}$ Mpc. Follow-up deep spectroscopy with the GTC telescope yields a redshift of 3353~$\pm$~29 km s$^{-1}$, making its association with the NGC~1211 galaxy (one of the most isolated galaxies known) highly likely. Both galaxies have compatible distances and redshifts, share a high peculiar velocity ($\sim$1000~km~s$^{-1}$) and lie within a void of radius {7~Mpc}. The local surface density is $\Sigma_5~\sim~0.06$ Mpc$^{-2}$, an order of magnitude smaller than the field population and similar to the voids found in the GAMA survey. Our work shows: i) The high potential of using optical ground-based photometry of associated globular clusters to explore distances to ultra-diffuse galaxies and ii) the presence of red ultra-diffuse galaxies even in the most sparse environments, suggesting a wide range of formation mechanisms.

\end{abstract}




\section{Introduction}
The faint galaxy population offers an extraordinary potential to unravel critical issues in the formation and hierarchical evolution of galaxies. For instance, the comparison of the number densities of observed satellite galaxies \citep[e.g.][and references therein]{2015ApJ...813..109D} with those in simulations \citep[e.g.][]{2008MNRAS.391.1685S,2008ApJ...679.1260M} has been a major test of the current $\Lambda$CDM cosmological paradigm \citep[e.g.][]{1999ApJ...524L..19M, 1999ApJ...522...82K,2018PhRvL.121u1302K}. This, and many other motivations, have triggered significant observational efforts to try to detect galaxies of increasingly low surface brightness. While the approach of counting stars is able to detect galaxies with surface brightnesses of up to $\mu$ $\approx$ 30 mag arcsec$^{-2}$ \citep[][]{2012AJ....144....4M, 2018ApJ...860...66M}, potentially offering a distance estimate through, e.g. the tip of the red giant branch \citep[][]{1993ApJ...417..553L}, this technique is limited by image resolution \citep[few Mpc for ground observations and $\sim$16 Mpc for the Hubble Space Telescope, see][]{2012MNRAS.421..190Z} and stochasticity \citep[][]{2010AdAst2010E..21W, 2014ApJ...787...19M}. This makes  integrated photometry the only alternative for  detecting sources beyond the resolution limit of  individual stars, and leads to greater uncertainties any distance measurements. 

Current deep photometric observations are able to detect unresolved and extremely low surface brightness galaxies (ultra-faint galaxies) in  nearby galaxy associations: e.g., M81 Group \citep{2009AJ....137.3009C,2013AJ....146..126C}, Sculptor Group \citep{2014ApJ...793L...7S}, Centaurus Group \citep{2014ApJ...795L..35C,2017A&A...597A...7M}, M101 Group \citep{2017ApJ...850..109B} and other nearby systems \citep[e.g.][]{2016A&A...588A..89J}. However, the presence of false projections in the line of sight limits this approach to very nearby structures where the size contrast against background sources is high.  


The existence of a subset of galaxies with very low surface brightness and large effective radius \citep[e.g.][]{1984AJ.....89..919S, 1988AJ.....96.1520F, 1988ApJ...330..634I, 1991ApJ...376..404B, 1997AJ....114..635D, 2003AJ....125...66C, 2015ApJ...807L...2K, 2015ApJ...809L..21M, 2015ApJ...813L..15M, 2017A&A...608A.142V, 2019MNRAS.tmp..241P}, recently coined as ultra-diffuse galaxies (UDGs) by \cite{2015ApJ...798L..45V}, has permitted the tracing of the abundance of the faint galaxy population at considerably greater distances by using spatial correlations with massive clusters of galaxies \citep[e.g.][]{2016A&A...590A..20V,2017MNRAS.468..703R,2017ApJ...844..157L}. Although adequate corrections for the presence of interlopers provides reasonable numbers on average densities \citep{2018MNRAS.481.4381M}, the distances of individual objects remain uncertain. Additionally, this selection based on spatial associations introduces a systematic bias towards high density environments. This could have an important impact on our understanding of these galaxies as the UDG population suffers from strong environmental variations in their average properties: They appear bluer and irregularly shaped in low density environments \citep[blue UDGs; ][]{2017MNRAS.468.4039R, 2017ApJ...836..191T, 2017MNRAS.467.3751B, 2017ApJ...842..133L, 2018A&A...614A..21J, 2018ApJ...857..104G, 2018ApJ...866..112G} in contrast to the red colours and roundish morphologies found in galaxy clusters (red UDGs). This suggests an evolutionary path similar to that followed by the general dwarf population: blue irregular to red spheroidal through environmental processing \citep[e.g.][]{2006MNRAS.366....2W, 2011ApJ...726...98K, 2015MNRAS.453...14Y}.

Therefore, the number densities of extended and low surface brightness galaxies in different environments offers very valuable information. However, even though these objects are readily  detected in photometric campaigns, spectroscopic observations are  expensive in terms of telescope time, and in some cases not feasible for galaxies with $\mu >$ 26 mag/arcsec$^{2}$. Particularly problematic is the case for the red objects in low density environments, in which the absence of gas and their extreme low surface brightness, in contrast with the high HI abundance and higher surface brightness of the blue objects\footnote{The fact that blue-UDGs are brighter means that often they do not meet the criterion of $\mu _g$(0) $>$ 24 mag/arcsec$^2$, which has contributed to their exclusion in UDG catalogues.} \citep{2017A&A...601L..10P, 2018ApJ...855...28S}, leads to the only option of confirming their distances through deep spectroscopic observations of their stellar content \citep{2016AJ....151...96M}. Another uncertainty to be taken into account, even if spectroscopy is carried out, is the possible deviation of radial velocities from the Hubble Flow \citep{2019MNRAS.tmp..733T}. This makes obtaining distances for large samples of diffuse objects problematic with the current methodologies and instrumentation. It is therefore, mandatory for the community, to explore new ways to obtain distances to extremely low surface brightness galaxies \citep{2019arXiv190107575C,2019arXiv190107578C}.

Interestingly, faint dwarf galaxies do have a number of globular clusters \citep[e.g.][]{2009MNRAS.392..879G}. This has been exploited extensively for the particular case of UDGs \citep{2016ApJ...819L..20B, 2016ApJ...822L..31P, 2016ApJ...830...23B, 2016ApJ...828L...6V,2017ApJ...844L..11V,2018MNRAS.475.4235A,2018ApJ...856L..31T,2018ApJ...862...82L,2019MNRAS.484.4865P} offering an interesting tool to obtain their virial masses indirectly \citep{2016ApJ...819L..20B,2019arXiv190100900B}. In this work we explore a different approach to obtain the distance of a diffuse galaxy detected in the IAC Stripe82 Legacy Survey. We use as a prior the fact that the luminosity function of GCs (the GCLF, a standard candle) appears to be nearly universal \citep[][and references therein]{2012Ap&SS.341..195R}, making it an important distance indicator for these systems. This technique has been widely used for massive galaxies, offering very competitive distance measurements, similar to other redshift-independent distance methods \citep[e.g.][]{1968JRASC..62..367R,1992PASP..104..599J, 1997eds..proc..254W,1999ASPC..167..204T, 2003LNP...635..281R, 2010ApJ...717..603V}, allowing for complementary calibration of the Hubble Constant \citep[e.g.][]{1968ApJ...152L.149S, 2000ApJS..128..431F}. This technique could be very useful in the specific case of extremely low surface brightness galaxies, presenting an alternative to the requirement of (sometimes not feasible) spectroscopy, as it exclusively requires deep photometric data with an appropriate spatial resolution.

This work is structured as follows: We describe the object and its detected globular cluster system in Section 2. In Section 3 we perform the fitting of the GCLF, obtaining the distance to the object. In Section 4 we suggest the spatial association of the object with the NGC~1211 galaxy. In Section 5 we present a general discussion and comments about this technique. We adopt the following cosmology ($\Omega_m$ = 0.3, $\Omega_\Lambda$ = 0.7 and H$_0$ = 73 km s$^{-1}$ Mpc$^{-1}$). We use the AB-magnitude photometric system. All Photometric and derived values have been corrected from extinction as A$^c$(\textit{u}) = 0.281 mag, A$^c$(\textit{g}) = 0.219 mag, A$^c$(\textit{r}) = 0.152 mag, A$^c$(\textit{i}) = 0.113 mag and A$^c$(\textit{z}) = 0.084 mag \citep{2011ApJ...737..103S}.

\section{A diffuse galaxy and its globular cluster system}\label{sec:2}

\begin{figure*}
  \centering
   \includegraphics[width=1.0\textwidth]{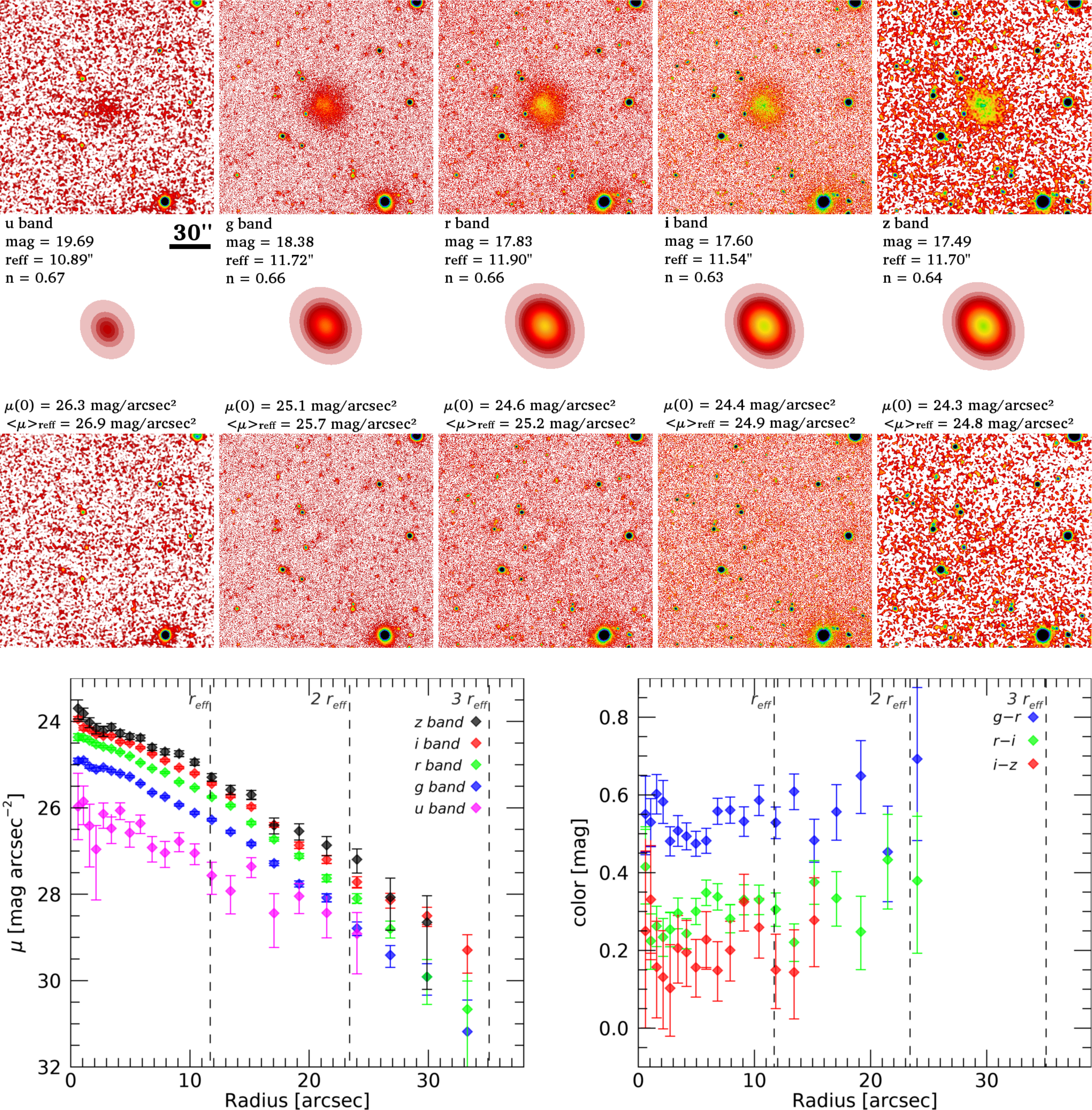}
    \caption{Structural and photometric properties of S82-DG-1. The top panels show the fitting of S82-DG-1 through a S\'ersic function. We show the original images of the galaxy (top row), models (middle row) and residuals (bottom row) in the five SDSS bands (columns). The text adjacent to the model shows the S82-DG-1 S\'ersic parameters and derived values. Original and residual images in \textit{u} and \textit{z} bands have been enhanced with a gaussian kernel smooth of 3 pixels. The stamps have a size of 2.6'$\times$2.6'. In the lower panels the photometric profiles of S82-DG-1 are shown. In the left lower panel, the photometric profiles are shown in individual bands. In the plot at the bottom right, the colour profiles are shown. Only points with enough statistical significance are included. Vertical dotted lines mark multiples of the effective radius {($r_{\mathrm{eff,g}}$ = 11.7'')}.}

   \label{fig:Panel_pho}
\end{figure*}

Using data from the IAC Stripe82 Legacy Survey \citep{2016MNRAS.456.1359F,2018RNAAS...2c.144R} we serendipitously discovered a galaxy with very low surface brightness and quite extended size at coordinates R.A. = +03$^{h}$ 07$^{m}$ 18.0$^{s}$, Dec. = $-$00\textdegree {} 47' 34.7''. There is no previous detection of this galaxy according to NASA/IPAC Extragalactic Database (NED). We name the object as S82-DG-1. We performed a photometric fit to the galaxy using the IMFIT \citep{2015ApJ...799..226E} package with a single S\'ersic model \citep{1963BAAA....6...41S}. The fittings were performed for all the \textit{u}, \textit{g}, \textit{r}, \textit{i} and \textit{z} individual bands, masking for external sources, using point spread function (PSF) deconvolution with the PSFs provided by the IAC Stripe82 Legacy Survey. We performed a previous fitting to the \textit{g}, \textit{r} and \textit{i} bands and we calculated the average ellipticity and position angle, producing $PA$ = 121.9 and $\epsilon$ = 0.15. Afterwards, we fix these values for all the bands, leaving the rest of parameters free. The results of the S\'ersic fitting are shown in  Fig. \ref{fig:Panel_pho}; upper panels. As can be seen, the S\'ersic model well approximates the morphology of the galaxy. We found good compatibility of the structural parameters between different bands, but a lower effective radius in the \textit{u} band, likely related to the extreme low surface brightness of S82-DG-1 in this spectral range together with the shallower \textit{u} band \citep[see a discussion of the effects of low signal to noise on the structural parameters of diffuse galaxies by][]{2018MNRAS.478..667P}. After analyzing the residual images, we found small deviations from the model evidenced by areas of very slight over-subtraction and asymmetries. We also created photometric profiles, this is shown in Fig. \ref{fig:Panel_pho}; lower panels. The profiles are compatible with a low-index S\'ersic profile, although slight deviations are found in the inner region within 1 effective radius, as in the case of residual images. Additionally, the colour profiles show some trend to redder colours at larger apertures. It is worth mentioning about the great depth limit of the profiles, which reach up to 3 effective radii in the case of individual bands (\textit{g}, \textit{r} and \textit{i}) and 2 effective radii for the colour profiles (\textit{g-r} and \textit{r-i}). Both the morphological and photometric parameters indicate that the galaxy is a low surface brightness quenched spheroid: $r_{\mathrm{eff,g}}$~=~11.7'', $n_g$~=~0.66, $\mu _g$(0)~=~25.1~mag/arcsec$^2$, $<\mu _g>$~=~25.7 mag/arcsec$^2$ and \textit{g-i}~=~0.78~mag.

\begin{figure}
  \centering
   \includegraphics[width=\columnwidth]{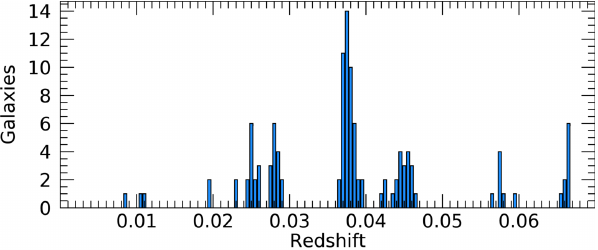}
    \caption{Redshift distribution of galaxies within a cone of 30 arcmin around S82-DG-1. The data was obtained from the SDSS DR14 Skyserver query.}
   \label{fig:Histogram}
\end{figure}

We searched for possible host structures with which to associate S82-DG-1 spatially. We did not find any Abell cluster of galaxies according to \cite{1991ApJS...77..363S}, however, using the SDSS spectroscopy in this area of the sky (a cone of 30' radius centered in the object; see Fig. \ref{fig:Histogram}), different galactic associations are found. A structure located at $z$~=~0.038 with 47 spectroscopic objects is the dominant structure in the line of sight. Other galactic associations with $\sim$15 spectroscopic objects are also found at $z$~=~0.025, 0.027 and 0.045, that can be considered groups of galaxies. Interestingly, if S82-DG-1 were located in any of these galactic associations, its effective radius would be considerable. However, since there are various structures along line of sight, it is not possible to confirm which might contain S82-DG-1, leaving its distance, and therefore its physical size and luminosity, uncertain.

Encouraged by the presence of numerous point sources clumped around the object after visual inspection, and the possibility to characterize its globular cluster system, we carried out a detection of GCs candidates in a wide area around the galaxy. Sources were detected and aperture photometry was performed on the original images of S82-DG-1 using Source Extractor \citep[SExtractor; ][]{1996A&AS..117..393B}. The detection was performed on the individual, un-sharp masked \textit{g}, \textit{r} and \textit{i} images. We ruled out the use of the \textit{u} and \textit{z} bands due to their shallowness after previous testing. Colours and magnitudes were determined using a 4-pixel diameter fixed aperture, and aperture corrections were applied by constructing a curve-of-growth with apertures ranging from 4 to 50 pixels. Point sources were identified  as objects with a \textit{g}-band SExtractor parameters \texttt{FLUX\_RADIUS} < 5.0 and \texttt{ELLIP} < 0.5. The final magnitudes were extinction corrected and we selected the GCs candidates as objects with 0.0~$<$~\textit{g-r}~$<$~1.3, 0.5~$<$ \textit{g-i}~$<$~1.5, 0.0~$<$~\textit{r-i} $<$~0.7. These criteria aim for a conservative selection of possible GCs taking into account the relatively large expected photometric uncertainties, rejecting likely background galaxies. The selected sample shows a significant  overdensity of GCs candidates, centered on the location of the galaxy, within a radius approximately twice of its physical size (Fig. \ref{fig:GCs}, upper left panel). This suggests the presence of a well-detected globular cluster system associated with the galaxy. However, this selection also shows the presence of interlopers as a background distribution of GC-like candidates in addition to the overdensity clumped around S82-DG-1. Fortunately, the area in which the galaxy is located contains a relatively low number of background galaxies, favouring a cleaner analysis. 

We obtained radial density profiles in circular annuli of the GCs candidates centered over the location of S82-DG-1, this is shown in Fig. \ref{fig:GCs}, upper right panel. We obtain a strong peak within a radius of 30'' ($\times$ 2.5 the effective radius of S82-DG-1: { $r_{\mathrm{eff}}$ = 11.7''}) with a maximum value of 57 GCs candidates per arcmin$^2$. This fact is in general agreement within uncertainties with the galactic effective radius vs. globular cluster system effective radius relation for ultra-diffuse galaxies by \cite{2017MNRAS.472L.104F}, which is a further indication of the existence of a detected globular clusters system around S82-DG-1. Therefore, we selected globular clusters candidates within a 30" radius (2.5$\times r_{\mathrm{eff}}$) for the GCLF analysis, resulting in 11 selected sources. We list their coordinates and photometry in Table \ref{table:gcs}.

\begin{table*}
\centering
\caption{Globular cluster candidates selected for the GCLF analysis.}
\begin{tabular}{cccccc}
\label{table:gcs} 
    ID & R.A. (J2000)  & Dec. (J2000)  & g & r & i  \\ 
      &[deg]&[deg]&  [mag]    & [mag]     & [mag] \\   
   \hline
GC1  & 46.82492 & $-$0.78917 & 24.15 $\pm$ 0.19 & 23.39 $\pm$ 0.08 & 23.22 $\pm$ 0.06\\
GC2  & 46.83103 & $-$0.79061 & 25.09 $\pm$ 0.39 & 24.97 $\pm$ 0.28 & 24.31 $\pm$ 0.14\\
GC3  & 46.82339 & $-$0.79130 & 24.42 $\pm$ 0.23 & 23.84 $\pm$ 0.11 & 23.69 $\pm$ 0.08\\
GC4  & 46.82688 & $-$0.79246 & 23.97 $\pm$ 0.17 & 23.54 $\pm$ 0.09 & 23.36 $\pm$ 0.07\\
GC5  & 46.82553 & $-$0.79315 & 23.79 $\pm$ 0.15 & 23.16 $\pm$ 0.07 & 22.79 $\pm$ 0.04\\
GC6  & 46.82814 & $-$0.79361 & 24.02 $\pm$ 0.18 & 23.56 $\pm$ 0.09 & 23.08 $\pm$ 0.05\\
GC7  & 46.82163 & $-$0.79442 & 25.07 $\pm$ 0.39 & 24.39 $\pm$ 0.17 & 23.85 $\pm$ 0.10\\
GC8  & 46.82393 & $-$0.79444 & 24.96 $\pm$ 0.35 & 23.98 $\pm$ 0.13 & 23.70 $\pm$ 0.08\\
GC9  & 46.82755 & $-$0.79442 & 24.87 $\pm$ 0.33 & 23.89 $\pm$ 0.12 & 23.86 $\pm$ 0.10\\
GC10 & 46.81908 & $-$0.79724 & 25.11 $\pm$ 0.40 & 23.85 $\pm$ 0.12 & 23.82 $\pm$ 0.09\\
GC11 & 46.82107 & $-$0.79936 & 25.11 $\pm$ 0.40 & 24.91 $\pm$ 0.27 & 24.30 $\pm$ 0.14\\

    \hline
\end{tabular}
\end{table*}

\section{Obtaining the distance through the GCLF}\label{sec:GCLF}

The most common way to obtain a distance value from the GCLF is to fit a gaussian function, whose peak or turn-off is considered a universal value, as introduced by \cite{1977MNRAS.180..309H}. Hence, the comparison of this peak with reference GCs distributions in local well-known galaxies allows the determination of the distance modulus. While this peak is considered nearly universal, the width of the distribution is a free parameter even though well correlated with the luminosity of the galaxy \citep[see e.g.][]{2007ApJS..171..101J}. For the specific case of metal-poor GCs in early-type galaxies, the peak of this distribution shows a low intrinsic dispersion between different galaxies \citep{2012Ap&SS.341..195R}. Thus, the analysis of the GCLF allows obtaining quite accurate redshift-independent distances, frequently applied to local massive galaxies, obtaining complementary constraints on Hubble's law \citep[see very first examples using the M87 galaxy by][]{1968JRASC..62..367R,1968ApJ...152L.149S,1985AJ.....90..595V}.

\begin{figure*}
  \centering
   \includegraphics[width=0.78\textwidth]{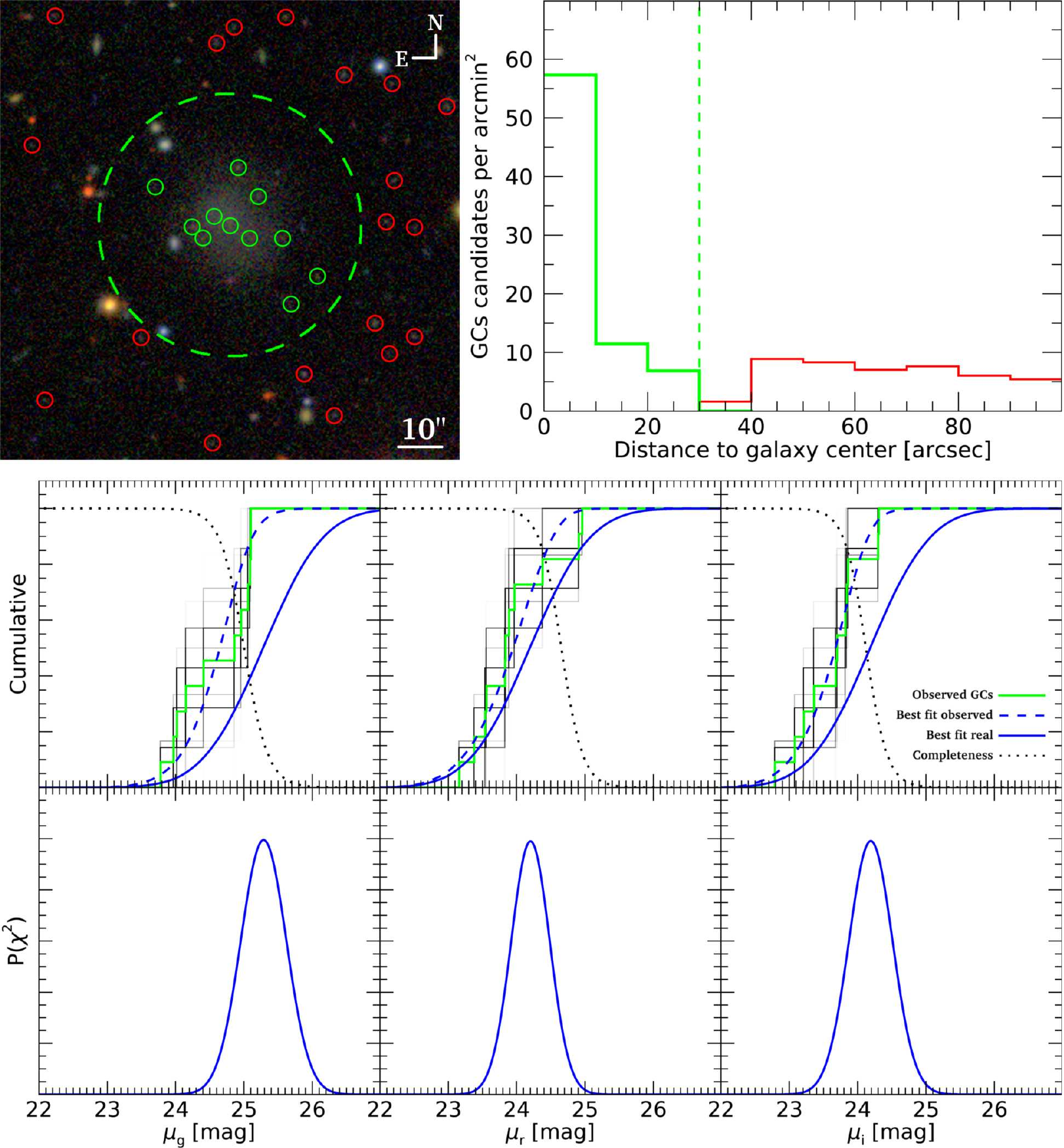}
    \caption{Analysis of the globular cluster system of S82-DG-1. Upper left panel: The sources for the GCLF analysis are selected within a radius of 2.5$\times r_{\mathrm{eff}}$ or 30'' (green dashed line), marked in green circles. The sources considered interlopers are marked with red circles. The image shows a zoom-in of the whole explored area, focusing on the GCs candidates around S82-DG-1. Upper right panel: Radial density profiles of the GCs (green) and interlopers (red).
    Bottom panel: The upper row shows the cumulative histograms of the GCs candidates (green). The black histograms correspond to different iterations in which 4.1 sources were extracted on average, plotted with a certain transparency. The blue line shows the best model and the dashed blue line shows the best model convolved with the completeness curve. Photometric completeness curves for point sources were calculated by injecting artificial PSFs into the original images and determining the recovered fraction with \texttt{SExtractor}, and are shown by dotted lines. The photometry is 100\% complete to $g = 24.1$, $r = 23.8$, $i = 23.2$ and 50\% complete to $g = 25.0$, $r = 24.7$, $i = 24.1$. The bottom row shows the probability distributions of the GCLF peak resulting from the $\chi^2$ minimization}

   \label{fig:GCs}
\end{figure*}

In this work, we aim to explore the possibility of obtaining distances to diffuse objects through the GCLF. However, the fact that we are using ground-based observations (in contrast to the usual space-based observations), the low mass of the diffuse objects (therefore, low number of GCs expected to be found) and the few studies about the globular cluster systems in such low mass galaxies with which to support this work, makes this task more challenging than usual GCLF characterizations. These circumstances create methodological difficulties that are summarized as follows: 

\begin{itemize}
    \item The seeing limited resolution together with the colour degeneracy between GCs and background point-like sources causes a relatively high fraction of interlopers. This uncertainty has to be taken into account for the GCLF fitting. 
    
    \item There is no complete detection of the GCLF. In Fig. \ref{fig:GCs} we show the magnitude cumulative histogram of the selected GCs candidates. As can be seen, the depth of the data limits the detection. Clearly, the fitting of the GCLF is blind beyond the region of completeness. This creates the need to fix the gaussian width of the GCLF ($\sigma_{GCLF}$) for a correct fitting. Additionally, this creates an important Malmquist bias in which the peak of the observed luminosity function may differ significantly from the real peak (see Appendix \ref{sec:App_sim}).
    
    \item The $\sigma_{GCLF}$ is a function of the luminosity of the galaxy, and as the distance of our object is unknown, this causes a degeneracy of not being able to know a priori the luminosity of the galaxy. However, this variation of $\sigma_{GCLF}$ is relatively small, reaching $\sigma_{GCLF}$ = 0.5 mag for the less massive galaxies (M$_B$ = $-$15) \citep[see e.g.][]{2007ApJS..171..101J}, being also approximately constant between different bands at a given luminosity.
    
    \item There are few studies on the detailed properties of the GCLF for diffuse galaxies. Work by \cite{2016ApJ...822L..31P} on the DF17 galaxy (8.4 $\times$ 10$^{7}$ M$_{\sun}$) and \cite{2019MNRAS.tmp..733T} on the KKSG04 galaxy (6.0 $\times$ 10$^{7}$ M$_{\sun}$) show $\sigma_{GCLF}$ $\sim$ 0.7 mag, a value in agreement with the $\sigma_{GCLF}$ vs. luminosity relation followed by the general dwarf population of similar luminosity \citep[e.g.][]{2010ApJ...717..603V}.

\end{itemize}

Taking into account everything discussed above, we proceed to obtain the peak of the GCLF in each band. Due to the low number of GCs candidates we use an approach based on cumulative histograms. In Fig. \ref{fig:GCs} we plotted the cumulative curve of the 11 GCs candidates associated with S82-DG-1 (green histogram). This cumulative curve, which approximates the S82-DG-1 GCLF, has two main uncertainties. On the one hand a certain number of sources are expected to be interlopers. Taking the average value of the density profiles between 30'' and 100'' (considered interlopers) we obtain 4.1 expected interlopers inside the 2.5$\times r_{\mathrm{eff}}$ of S82-DG-1. On the other hand, the GCLF suffers from photometric incompleteness. We calculate the incompleteness curve by injecting/recovering PSFs in each band, and this is shown as a dotted curve in each band in Fig. \ref{fig:GCs}. This incompleteness generates a strong bias in which the observed GCLF tends to be brighter than the real GCLF. We show a visualization of this effect in Appendix \ref{sec:App_sim}. To take these uncertainties into account, we obtain the peak of the luminosity function through a $\chi^2$ minimization:
$$\chi^2(\mu) = \sum_{m}  \left[\frac{GCLF'_{m}-LF_{m}(\mu)}{\epsilon_{m}}\right]^{2}  $$
where $GCLF'_{m}$ is the cumulative histogram of the 11 GCs to which we randomly extract a number of 4.1 sources (4 or 5 sources in each iteration with a probability of 4.1 sources). $LF_{m}(\mu)$ is the cumulative histogram of the comparison model centered on $\mu$, convolved with the completeness curve, or in other words, the cumulative curve of observable GCs of a gaussian function with peak at $\mu$. $\epsilon_{m}$ are the uncertainties in the cumulative curves in each magnitude bin $m$, and were calculated by injecting luminosity functions with a number of sources similar to this case and taking into account the photometric uncertainties of the GCs in the magnitude bin $m$. Since both the $\sigma_{GCLF}$ and the number of GCs of S82-DG-1 are unknown, we set a range of these parameters in the model. We set an arbitrary $\sigma_{GCLF}$ in each iteration within a range $\sigma_{GCLF}$ = [0.5,1.1] mag, motivated by the dispersion of this parameter observed in galaxies of similar luminosity \citep{2007ApJS..171..101J,2010ApJ...717..603V}. Additionally we limited the number of GCs of the model. After preliminary tests we observed a high degeneracy between the number of GCs, $\sigma_{GCLF}$ and $\mu$. This causes that the peak of the distribution $\mu$ can take arbitrarily faint values whenever $\sigma_{GCLF}$ and the number of GCs is sufficiently high. This would make that luminosity functions with a very high $\sigma_{GCLF}$, an arbitrarily high number of GCs (thousands of sources) and a very faint peak ($\mu$) could reproduce the observed properties of the GCLF of S82-DG-1 (only the few brightest GCs of the thousands of the GCLF would be observed). However, we consider this scenario as highly unlikely and incompatible with observations. For this we set the maximum number of GCs of the model to 50 GCs, motivated observationally for galaxies of this luminosity range \citep[see Fig. 5 by][]{2019MNRAS.484.4865P}. {Only two UDGs in the Coma cluster are thought to have more than 50 GCs \citep[DFX1 63 $\pm$ 17 GCs and DF44 76 $\pm$ 18 GCs; ][]{2017ApJ...844L..11V}}. The optimal case would be to use an arbitrarily high number of GCs in the model to avoid the shot noise, having a continuous cumulative histogram for the model. Hence, assuming 50 GCs instead of an arbitrarily high number, will increase the $\chi^2$ dispersion in each iteration. Nevertheless, with a sufficiently high number of iterations this effect is minimized. Additionally, we tested possible biases  by assuming a smaller number of GCs. We did not find any difference in the most likely recovered peak, being the only bias the one discussed above for an arbitrarily high number of GCs. This leads us to rely on 50 GCs for the model as the best compromise between reducing the $\chi^2$ shot noise and avoiding a bias towards scenarios not compatible with the observations. Note that the errors ($\epsilon_m$) are based on this assumption. Summarizing: we perform iterations in which in each step we extract a number of sources to the observed S82-DG-1 GCLF, obtaining a vector $\chi^2(\mu)$ by comparison to a luminosity function model located in different peaks ($\mu$) with the properties described above. The result after 20,000 iterations is plotted in Fig. \ref{fig:GCs}. We show the different S82-DG-1 GCLF cumulative curves in which 4.1 sources were extracted on average as the black curves with a certain transparency. It allows to visualize the different steps of each iteration (darker areas are more frequent/likely histograms). We plot the average probability density distribution $P(\chi^2)$ for all the iterations performed (Fig. \ref{fig:GCs}, bottom rows). As can be seen, it has a well defined Gaussian shape. Therefore we fit the probability distribution with a Gaussian function, considering the center of the fitted distribution as the most likely peak and the width its error (1$\sigma$). Additionally, we plot the cumulative curve of the most likely luminosity function, both for the real case (without completeness effects, continuous blue curve) and for what would be observed under the incompleteness regime of the dataset (dashed blue curve). As can be seen, the best fit under incompleteness conditions is located in the average of the different $GCLF'_{m}$ iterations (interlopers subtractions). Finally, the values obtained through this method in the different bands are: $\mu_g$ = 25.27 $\pm$ 0.33 mag, $\mu_r$ = 24.20 $\pm$ 0.28 mag and $\mu_i$ = 24.19 $\pm$ 0.32 mag.

As reference for the GCLF peak we use the value reported by \cite{2012Ap&SS.341..195R} of $\mu_{V,ref}$~=~$-$7.66 mag. We transformed this value into SDSS bands using stellar population models by \cite{2015MNRAS.449.1177V} with a single stellar population model of 6 Gyr and [Fe/H]~=~$-$1 (see next section). The transformed values are $\mu_{g,ref}$~=~$-$7.27 mag, $\mu_{r,ref}$~=~$-$7.86 mag and $\mu_{i,ref}$~=~$-$8.13 mag, in agreement to those values reported by \cite{2007ApJS..171..101J, 2010ApJ...717..603V} ($\mu_{g}$~$\approx$~-7.20 mag) for galaxies of similar luminosity. We also assume an error from the intrinsic dispersion among galaxies for the GCLF peak of $\Delta \mu_{j,ref}$~=~$\pm$~0.2 mag, which is a more conservative value than the one provided by \cite{2012Ap&SS.341..195R} of $\Delta \mu_{j}$~=~$\pm$ 0.09 mag, trying to take into account the uncertainty due the few studies of GCLFs in low mass galaxies and the uncertainty in the globular cluster system metallicity \citep{1995AJ....110.1164A}. Therefore, the value of the distance modulus for each j-th band using the globular cluster system and its error are:
$$ (m-M)_{j}=\mu_{j}-\mu_{j,ref} $$
$$ \Delta(m-M)_{j}=\Delta\mu_{j}+\Delta \mu_{j,ref} $$
The peak values are transformed into the following distance modules: (m-M)$_{g}$ = 32.54 $\pm$ 0.53 mag, (m-M)$_{r}$ = 32.06 $\pm$ 0.48 mag and (m-M)$_{i}$ = 32.32 $\pm$ 0.52 mag, values compatible between bands within the error intervals. As the final value for the distance modulus we calculate the weighted mean using the three available bands, resulting in (m-M) = 32.29 $\pm$ 0.29 mag, equivalent to D = 28.7$_{-3.6}^{+4.2}$ Mpc, which we consider the distance for S82-DG-1 through the GCLF analysis. 

As an additional test, we performed the same procedure but without taking into account the incompleteness effects, that is, a direct fitting of the observed S82-DG-1 GCLF. We obtained: $\mu_g$ = 24.62 $\pm$ 0.18 mag, $\mu_r$ = 23.88 $\pm$ 0.18 mag and $\mu_i$ = 23.57 $\pm$ 0.17 mag. The results of this test are significantly brighter peaks, which would be equivalent to a distance of D = 22.6$_{-2.2}^{+2.4}$ Mpc, significantly closer. This shows the great importance in taking into account the effects of incompleteness in the fit of the GCLF.

{Finally, assuming a fixed distance of D = 28.7 Mpc, $\sigma_{GCLF}$ = 0.7 mag and taking into account the presence of interlopers, the most likely number of GCs for S82-DG-1 is 14.6 $\pm$ 4.8 GCs, assuming as error the standard deviation between bands.}

\section{Spatial association with the NGC~1211 galaxy and spectroscopic confirmation}

The analysis of the GCs system associated with S82-DG-1 provides a redshift-independent distance of D~=~28.7$_{-3.6}^{+4.2}$ Mpc. We explored, using the NED Database, nearby galaxies or galactic associations with which to associate our object spatially. The result is the likely association with the NGC~1211 galaxy, located at only 6.4 arcmin of angular separation to the west of S82-DG-1. The NGC~1211 galaxy has three redshift-independent distances. \cite{2013ApJ...771...88L}, using the Tully-Fisher relation, obtained m-M~=~32.27~$\pm$~0.69 mag equivalent to D~=~28.4$_{-7.7}^{+10.6}$ Mpc, \cite{2014MNRAS.445.2677S}, using the Fundamental Plane method, obtained m-M~=~32.38~$\pm$~0.51 mag equivalent to D~=~29.9$_{-6.2}^{+7.9}$ Mpc and \texttt{Cosmicflows-3} \cite[][]{2016AJ....152...50T} gives a value of m-M~=~32.29~$\pm$~0.50 mag equivalent to D~=~28.7$_{-5.9}^{+7.4}$ Mpc. These values overlap with the distance obtained through the GCLF analysis, suggesting the association between S82-DG-1 and the NGC~1211 galaxy. Further, the NGC~1211 galaxy exhibits very distinctive and interesting properties. According to the work by \cite{2012AJ....144...57F}, NGC~1211 is considered an "extremely isolated galaxy". The authors define such a galaxy as "\textit{isolated from nearest neighbors more luminous than M$_V$~=~$-$~16.5 mag by a minimum distance corresponding to 2.5 Mpc and 350 km s$^{-1}$ in redshift space}". Another interesting characteristic of NGC~1211 is its high peculiar velocity, which is perhaps related to  its relative isolation. While the redshift-independent values for its distance indicate D~=~29.1$_{-4.0}^{+4.6}$ Mpc (value obtained averaging the available distances for NGC~1211), its radial velocity obtained from SDSS spectroscopy is V$_{helio}$~=~3211 $\pm$~3~km~s$^{-1}$ (z~=~0.01071; see Fig. \ref{fig:Histogram}). Once corrected to the Cosmic Microwave Background reference (V$_{CMB}$~=~3015~$\pm$~3~km~s$^{-1}$), NGC~1211 has a peculiar velocity of V$_{pec,CMB}$~=~891$_{-336}^{+292}$ km s$^{-1}$, a rather high value. Note that the naive distance obtained directly from the redshift would be equivalent to 41.6 Mpc, and hence incompatible with the available redshift-independent distances. This high peculiar velocity gives us a very interesting opportunity to test if S82-DG-1 is associated with the NGC~1211 galaxy, and in the process, corroborate the effectiveness of our distance method through the GCLF analysis, as if in fact they are associated, the radial velocity of S82-DG-1 should have a similar high peculiar velocity.

In order to confirm the spatial association with the NGC~1211 galaxy and the reliability of our distance estimation based on the GCs, we obtained deep spectroscopy of S82-DG-1 with the 10.4m Gran Telescopio Canarias (GTC) telescope in the Roque de los Muchachos Observatory, La Palma, Spain. The observations were carried out through Director Discretionary Time (GTC06-17BDDT program: \textit{"Detection of an ultra-diffuse galaxy in a low density environment using as prior distance its globular cluster system. Spectroscopic confirmation"}) on the 18th and 19th of January, 2018. We used the OSIRIS spectrograph with the R2000B grism, covering the spectral range 3950 - 5700 \AA. This configuration provides a spectral resolution of $\sim$9.5 \AA, equivalent to $\sim$~300~km~s$^{-1}$. The slit width was of 2.5'' to maximize the light gathered by the instrument due to the extremely low surface brightness of the target. The sky conditions were good, dark time and $\sim$2'' seeing. The total exposure time was 16800 sec (4.66 h), divided in 12 individual exposures of 1400 sec. The reduction process was carried out with an automatic pipeline that includes the standard steps of bias subtraction and flat-fielding, $\lambda$ calibration,  sky subtraction using Kelson algorithm \citep{2003PASP..115..688K} and coadding of the individual exposures. The two-dimensional spectrum was collapsed in a range of 60 pixels (15'') that covers the region of the 65 percentile in flux along the slit. The final stacked one-dimensional spectrum presented some residual wiggles due to sky variations increased due to the extremely low surface brightness of the object ($<\mu _g>$~=~25.7~mag~arcsec$^{-2}$). This was corrected by fitting a low order polynomial in a similar way to that performed by \cite{2018A&A...617A..18R} observing a target of similar surface brightness with a similar instrumental configuration of the OSIRIS spectrograph. In spite of this correction, the regions of the spectrum outside the range 4050 - 4450 {\AA} still contains some residual fluctuations, and a lower signal to noise in the redder part of the spectra, likely associated to the higher sky emission in this spectral range and exacerbated by defects in the OSIRIS CCD. Therefore, we restrict our subsequent analysis to the more reliable region between 4050 - 4450 {\AA}. We consider it worth commenting on the  difficulties of spectroscopic observations for objects of such low surface brightness, whose brightness is up to two orders of magnitude lower than the sky brightness by itself.

\begin{figure}
  \centering
   \includegraphics[width=\columnwidth]{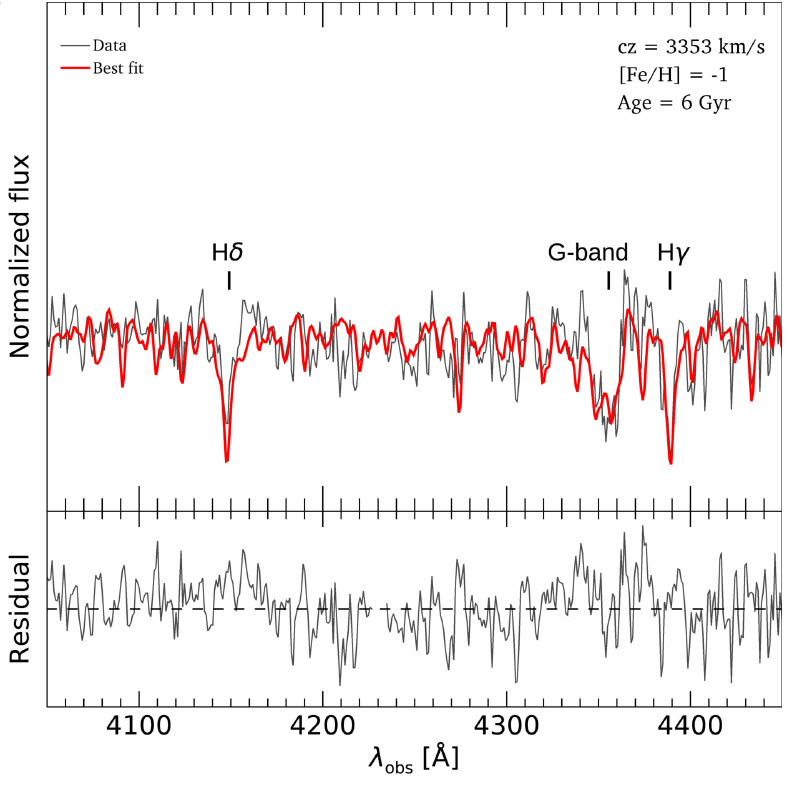}
    \caption{Final stacked spectrum (4.6 hours) using the 10.4m GTC telescope of S82-DG-1. The spectrum and residual are shown with the grey line. The indicated values of radial velocity, metallicity and age correspond to the best fit and its model spectrum is showed with the red line.}

   \label{fig:Spectrum}
\end{figure}

The final spectrum is plotted in Fig. \ref{fig:Spectrum}. The spectrum has signal to noise of around 10 per pixel (1 pix = 0.9 \AA) in the H$\delta$ region. To obtain its radial velocity and a possible stellar population analysis, we firstly tried a full spectrum fitting using the pPXF software \citep{2017MNRAS.466..798C}. However, given the low spectral resolution of $\sim$300 km s$^{-1}$ and the narrow spectral range of 4050 - 4450 {\AA}, the obtained model is not reliable enough. These circumstances also do not allow us to constrain any stellar population properties through STARLIGHT \citep{2005MNRAS.358..363C} or STECKMAP \citep{2006MNRAS.365...74O}. Therefore, we compared visually the spectrum with synthetic single stellar population galactic models from the MILES library \citep{2010MNRAS.404.1639V,2011A&A...532A..95F}. This visual comparison leads us to limit the age and metallicity of our object in a region around [Fe/H]~=~$-$1 and Age~=~6~Gyr. We obtained radial velocities through cross-correlation \citep{1979AJ.....84.1511T} between a set of synthetic spectra and our object spectrum using the \textit{fxcor} IRAF routine. We tested the cross-correlation with ages and metallicities far from the mentioned most likely value and we obtained the lowest error with these values in particular. 

\begin{figure}
  \centering
   \includegraphics[width=\columnwidth]{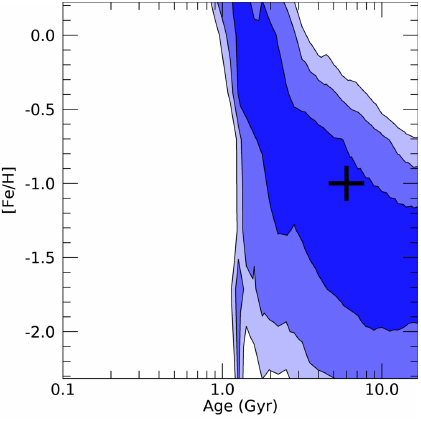}
    \caption{Age vs. metallicity probability contours for S82-DG-1 derived from the available photometric data. The blue areas show the 1, 2 and 3$\sigma$ probability contours. The black cross indicates the value of age and metallicity obtained from the spectrum.}
   \label{fig:Pops}
\end{figure}

As a further test, we compared the probability map of age vs. metallicity derived from the photometric data using the available \textit{u}, \textit{g}, \textit{r}, \textit{i} and \textit{z} bands through comparison with photometric predictions by \cite{2015MNRAS.449.1177V}. As can be seen in Fig. \ref{fig:Pops}, the values from the spectra are located within the 1$\sigma$ contour, i.e. compatible with the photometric data, giving some robustness to this value. Therefore, we consider [Fe/H]~=~$-$1 and Age~=~6 Gyr as the best fit model, {being in general agreement with galaxies of similar characteristics \citep[e.g.][]{2017ApJ...838L..21K, 2018ApJ...859...37G, 2018ApJ...858...29P, 2018MNRAS.479.4891F,2018MNRAS.478.2034R, 2018arXiv181207346F}, but see work by \cite{2019MNRAS.484.3425M} in the DGSAT-I galaxy}. We find a final radial velocity for S82-DG-1, once corrected to the heliocentric frame of V$_{helio}$~=~3353~$\pm$~29 km s$^{-1}$ (z~=~0.01118~$\pm$~0.00010).

\begin{figure*}
  \centering
   \includegraphics[width=\textwidth]{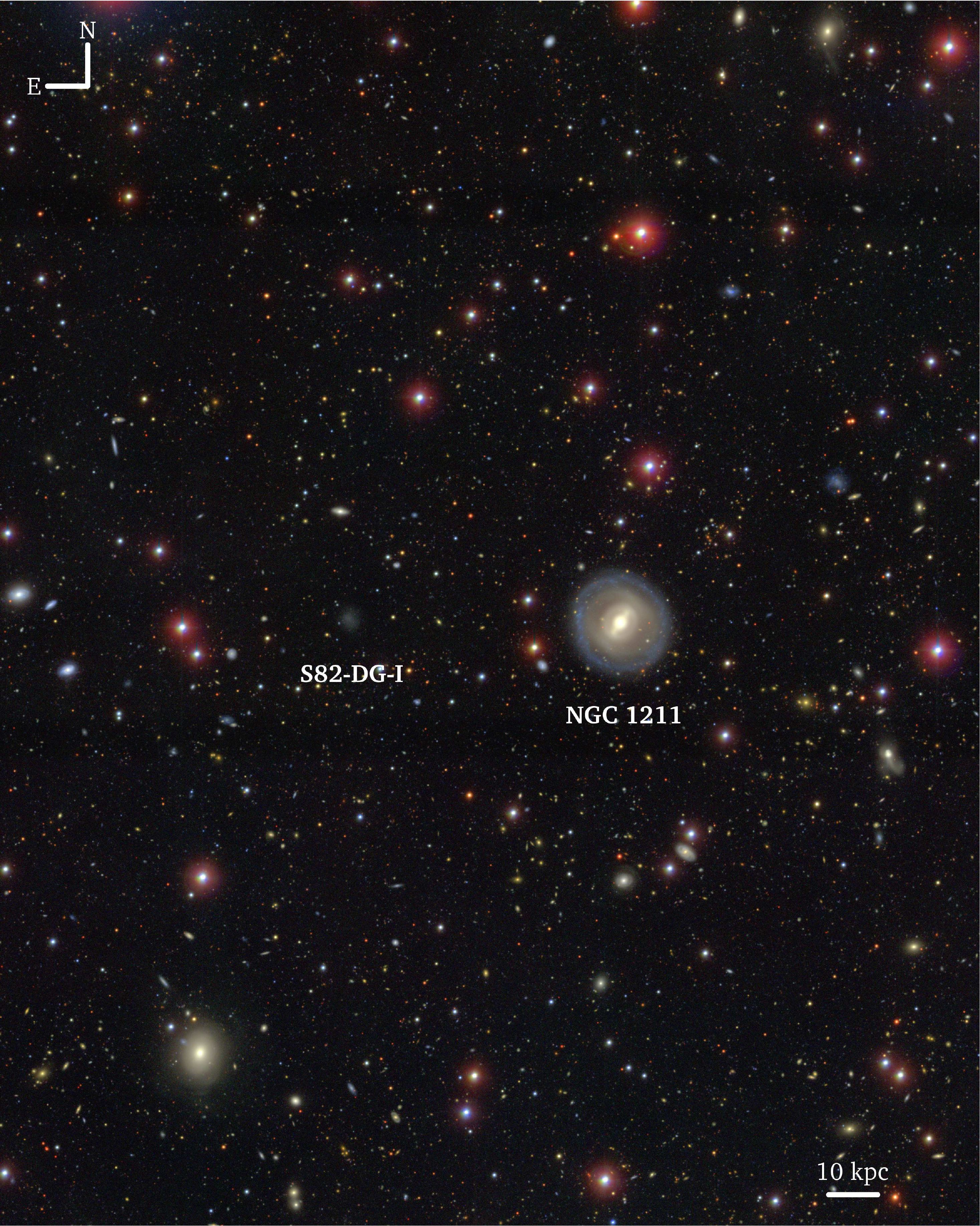}
    \caption{Composite colour image in \textit{g}, \textit{r} and \textit{i} SDSS bands from the IAC Stripe82 Legacy Survey showing the environment (23' $\times$ 29') of the S82-DG-1 and NGC~1211 galaxies. The scale marking 10 kpc ($\sim$70'') corresponds to the distance of 29 Mpc, and it is only valid for the S82-DG-1 and NGC~1211 galaxies. This work has demonstrated that both galaxies are spatially associated through a method in which globular clusters are used as distance estimators.}

   \label{fig:UDG_NGC1211}
\end{figure*}

Trying to address in a more precise way the association of S82-DG-1 with NGC 1211 once known its redshift, we used the NED database to select galaxies brighter than 16 mag in the \textit{r} band in a 2$\times$2 degree area around S82-DG-1 (approximately 1 Mpc in projection assuming a distance of 28.7 Mpc for S82-DG-1) and with radial velocity of $\pm$ 1000 km s$^{-1}$ from that obtained for S82-DG-1. Galaxies with this criterion are shown in table \ref{tab:Candidates}. Of all the galaxies capable of hosting S82-DG-1, based exclusively on the redshift, only NGC 1211 has a radial velocity compatible with S82-DG-1, having also a compatible redshift-independent distance. It makes the association between NGC 1211 and S82-DG-1 highly likely and gives reliability to the GCLF analysis method. We show in Fig. \ref{fig:UDG_NGC1211} a colour composite image of the field surrounding the NGC~1211 and S82-DG-1 galaxies. In Table \ref{tab:Distances} we summarize the distance values available for both objects.

\begin{table}
\centering
\caption{Galaxies candidates to be associated with S82-DG-1 (see text). Distances (if available) refer to redshift-independent methods.}
\normalsize
\begin{tabular}{lccc}
\label{tab:Candidates} 
Galaxy  & V $-$ V$_{S82-DG-1}$ & Distance & \textit{r} mag \\
 &  [km s$^{-1}$] & [Mpc] & [mag]\\
\hline
UGC 02482 & $-$712 & - & 15.5\\
UGC 02479 & $-$511 & 49.4$_{-4.3}^{+4.8}$ Mpc & 15.1\\
NGC~1211 & $-$142 & 29.1$_{-4.0}^{+4.6}$ Mpc & 12.2\\
UGC 02517  & 502 & - & 14.1\\
UGC 02505  & 717 & - & 15.7\\
NGC 1194  & 700 & - & 12.5\\
\hline
\end{tabular}
\end{table}

\section{Discussion}

In this paper we have explored the possibility of obtaining distances to diffuse galaxies through ground-based photometry of their globular cluster systems. For this we have carried out a thorough analysis of the GCLF of a diffuse galaxy detected in the IAC Stripe82 Legacy Survey, named S82-DG-1 in this work. As follow-up, we have also carried out deep spectroscopic observations of S82-DG-1 to confirm its distance and prove the reliability of the method. The analysis of the S82-DG-1 GCLF has allowed us to obtain a distance modulus of 32.29 $\pm$ 0.29 mag, equivalent to D = 28.7$_{-3.6}^{+4.2}$ Mpc. There are numerous lines of evidence suggesting the spatial association of S82-DG-1 with the NGC~1211 galaxy. Up to three different redshift-independent methods, the fundamental plane, the Tully-Fisher relation and \texttt{Cosmicflows-3} for NGC~1211\footnote{We do not perform a similar analysis for the NGC~1211 GCLF. The reason is the difficulty in extracting a precise photometry for GCs due to the morphology of NGC~1211, which has a remarkable central bar with three different rings one of which is clearly young with many knots of current star formation. This creates problems in the identification and photometry of GC candidates given our limited spatial resolution.} and the GCLF analysis for S82-DG-1 are in agreement. The fact that both galaxies have similar radial velocities, with the addition of the high peculiar velocities that both galaxies share, makes their spatial association highly likely.

\begin{table}
\centering
\caption{Summary of distances and radial velocities for S82-DG-1 and NGC~1211. The peculiar velocity is calculated by averaging the available distances (Tully-Fisher, fundamental plane and cosmicflows-3 methods) for NGC~1211 (D = 29.1$_{-4.0}^{+4.6}$ Mpc) and the GCLF distance for S82-DG-1 (D = 28.7$_{-3.6}^{+4.2}$ Mpc).}
\normalsize
\begin{tabular}{lll}
\label{tab:Distances} 
Method & S82-DG-1 & NGC~1211 \\
\hline
FP & - & 29.9$_{-6.2}^{+7.9}$ Mpc \\
TF  & - &  28.4$_{-7.7}^{+10.6}$ Mpc \\
\texttt{Cosmicflows-3} & - & 28.7$_{-5.9}^{+7.4}$ Mpc \\
GCLF &   28.7$_{-3.6}^{+4.2}$ Mpc & - \\
\hline
V$_{helio}$ & 3353 $\pm$ 29 km s$^{-1}$ & 3211 $\pm$ 3 km s$^{-1}$ \\
V$_{pec,CMB}$ & 1067$_{-307}^{+263}$ km s$^{-1}$ &  891$_{-336}^{+292}$ km s$^{-1}$  \\
    \hline
\end{tabular}
\end{table}

\begin{table}
\begin{center}
\caption{Summary of properties for the S82-DG-1 galaxy.}
\normalsize
\begin{tabular}{ll}
\label{tab:UDG_properties}
Parameter & Value \\
\hline
R.A. & +03$^{h}$ 07$^{m}$ 18.0$^{s}$\\
Dec. & $-$00\textdegree {} 47' 34.7''\\
\textit{u}-band & 19.69 $\pm$ 0.40 mag\\
\textit{g}-band & 18.38 $\pm$ 0.03 mag\\
\textit{r}-band & 17.83 $\pm$ 0.03 mag\\
\textit{i}-band & 17.60 $\pm$ 0.07 mag\\
\textit{z}-band & 17.49 $\pm$ 0.15 mag\\
{\textit{g-i}} & 0.78 $\pm$ 0.10 mag \\
$<\mu _g>$ & 25.7 $\pm$ 0.1 mag/arcsec$^2$ \\
$\mu _g$(0) & 25.1 $\pm$ 0.1 mag/arcsec$^2$ \\
S\'ersic index & 0.65 $\pm$ 0.01\\
Axis ratio & 0.85 $\pm$ 0.01 \\
{Position angle} & 122 $\pm$ 2 degrees\\
$[$Fe/H$]$ $^\dagger$ & $-$1 \\
Age $^\dagger$ & 6 Gyr \\
{M/L$_{r-band}$ $^\dagger$} & 1.3 $\Upsilon_{\sun}$ \\
Distance & 28.7$_{-3.6}^{+4.2}$ Mpc \\
Redshift & 0.01118 $\pm$ 0.00010 \\
Effective radius (\textit{g}-band) { }{ }{ }{ }{ }{ }{ }{ }{ }{ }{ }{ }{ }{ }{ }{ }{ }   & 1.6 $\pm$ 0.2 kpc  \\
M$_{g}$ & $-$13.9 $\pm$ 0.3 mag \\
{M$_{*}$  $^{\dagger \dagger}$} & 6.2 $_{-1.4}^{+2.0}$ $\times$ 10$^{7}M_{\sun}$\\
    \hline
\end{tabular}
\end{center}

$^\dagger$ Most likely values through cross-correlation with synthetic spectra. No error interval is available.

$^{\dagger \dagger}$ Value obtained using the mass to light ratio in the \textit{r}-band.

\end{table}

\begin{figure}
  \centering
   \includegraphics[width=\columnwidth]{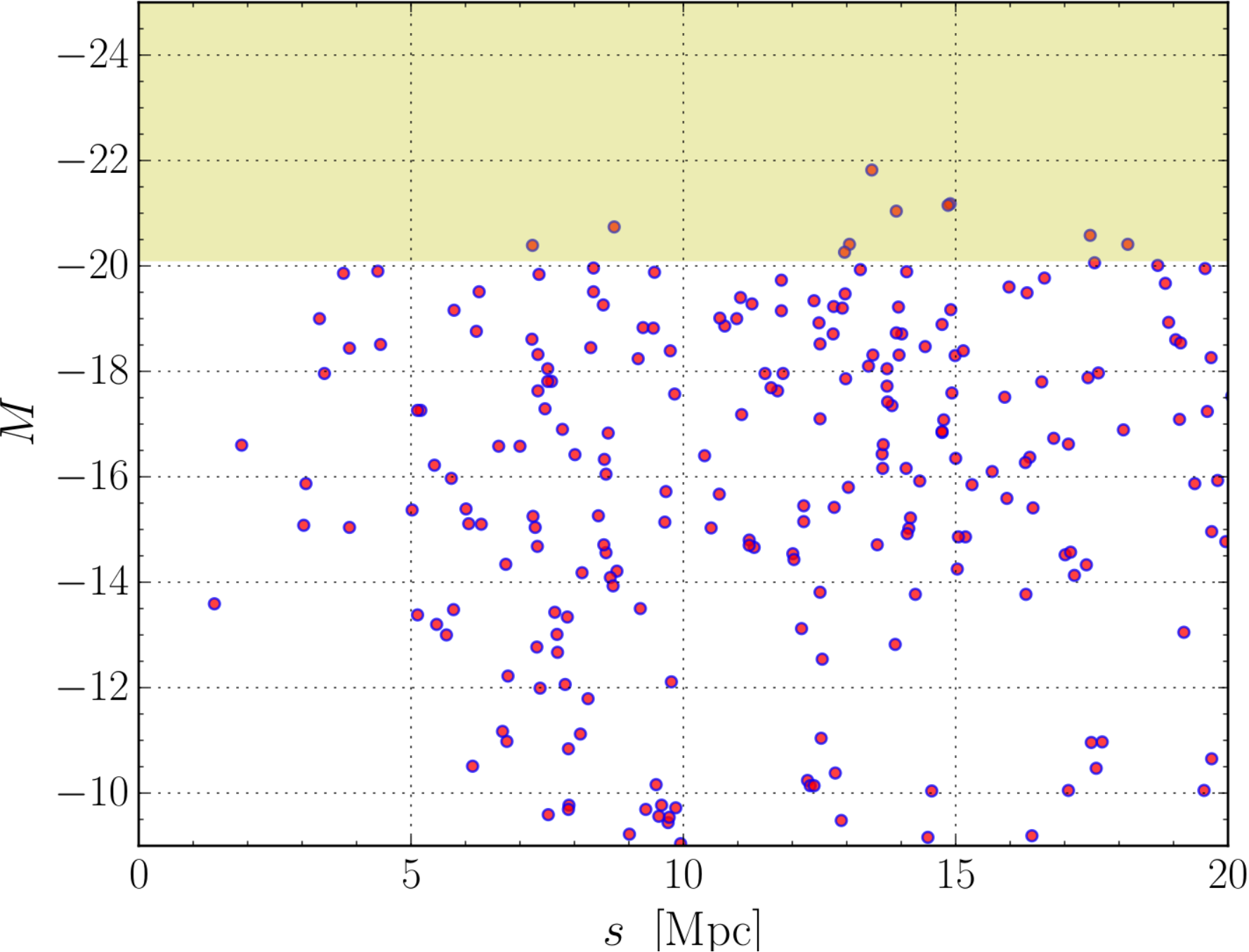}
   \caption{Distribution of galaxies within 10 degrees of S82-DG-1 and NGC~1211. Following the criteria adopted by the GAMA survey \citep{2015MNRAS.453.3519P}, galaxies brighter than $-$20.09 in the \textit{r} band define the properties of the void. The nearest galaxy lies at over 7 Mpc and shows that S82-DG-1 and NGC~1211 lie in a well-defined void.}
   \label{fig:Void}
\end{figure}

The calculated distance for S82-DG-1 gives an effective radius of 1.6 kpc, which together with its low surface brightness of $\mu _g$(0) = 25.1 makes this galaxy fall into the category of the so-called ultra-diffuse galaxies. The comprehensive  analysis of S82-DG-1 carried out in this work has allowed us to obtain a detailed characterization of this ultra-diffuse galaxy. We summarize its properties in Table \ref{tab:UDG_properties}. The environment of S82-DG-1, characterized indirectly with the work by \cite{2012AJ....144...57F}, shows that NGC~1211 and S82-DG-1 are located in an extremely low density environment. We have used the NED database to search for galaxies with redshifts within 10 degrees of this pair of galaxies, and compute the redshift-space distances as:
$$ s =  \frac{1}{H_0}  \sqrt{ V^2 + V_{NGC~1211}^2 - 2 V V_{NGC~1211} \cos\theta} $$
where the velocities are in the heliocentric frame, and $\theta$ is their angular separation. The resulting distribution of separations is given in Fig. \ref{fig:Void} where the absolute magnitudes are also indicated. The distribution of neighbouring galaxies confirms the isolation of this pair, and adopting the criterion of the GAMA survey for voids (no galaxy brighter than $M_r = -20.09$), the nearest galaxy would be at some 7 Mpc away. This is the typical size of voids found in the GAMA survey \citep{2015MNRAS.453.3519P} and similar to the ones found in other void surveys such as SDSS \citep{2012MNRAS.421..926P}. Using the fifth nearest neighbour as a proxy for the surface density, we get $\Sigma_5 = 0.06$ Mpc$^{-2}$, an order of magnitude smaller than the field population and confirms that this void has properties similar to the ones found in GAMA ( $\Sigma_5 = 0.09$ Mpc$^{-2}$). It makes S82-DG-1 the most isolated red-UDG so far.

While many examples of star-forming, irregularly shaped and large sized low surface brightness galaxies (blue-UDGs) are found mostly in the field \citep{2017MNRAS.468.4039R, 2017ApJ...836..191T, 2017MNRAS.467.3751B, 2017ApJ...842..133L, 2018A&A...614A..21J, 2018ApJ...857..104G, 2018ApJ...866..112G}, S82-DG-1 and DGSAT-I \citep{2016AJ....151...96M} are striking examples of red-UDGs whose formation does not require high density environments such as galaxy clusters. The fact of finding red or quenched UDGs in such low density environments promote the idea that these galaxies undergo a formation process not exclusively related to high density environments, where they are found in large numbers. In addition, the coexistence of red and blue UDGs in groups of galaxies are evidence of quenching and subsequent passive evolution prior to their infall to a galaxy cluster via aggregation of minor mergers \citep{2017MNRAS.468.4039R, 2018MNRAS.479.3308A}. 

The peculiar properties of S82-DG-1 make it an excellent laboratory for testing formation mechanisms of ultra-diffuse galaxies. The extremely low density environment of S82-DG-1 raises the question of what could be the cause of its quenching. It could indicate the result of internal processes, as the consequence of supernova feedback \citep{2017MNRAS.466L...1D}, which would support this formation mechanism for ultra-diffuse galaxies. Additionally, the exceptional depth of the photometric profiles made in Fig. \ref{fig:Panel_pho} allows us to observe a tendency to redness of the $\textit{g-r}$ and $\textit{r-i}$ colour profiles as the radius increases. This could indicate an older component to the main luminosity-weighted average colour, suggesting an extended star formation history, which is one of the predictions of this scenario and in agreement with observational results \citep{2018MNRAS.479.4891F,2018MNRAS.478.2034R}. However, the low signal to noise of the profiles makes this result uncertain and deeper observations are needed. It is also interesting to analyze the possible influence of NGC 1211 on the properties of S82-DG-1. In this aspect it is key to verify if S82-DG-1 really is a satellite of NGC~1211 or if it is only in projection proximity. A fact that would indicate that it is a true satellite is the presence of another object associated with the system: J030627.79-005247.9, located at only 66 kpc in projection from the NGC~1211 galaxy (assuming 29.1 Mpc as distance for the system). This galaxy is a star-forming dwarf (M$_{r}$ = $-$10.9 mag) with a radial velocity of $-$107 km s$^{-1}$ relative to that of NGC~1211 galaxy (38 km s$^{-1}$ from that of S82-DG-1). This would indicate that NGC~1211, despite its large scale isolation, has a satellite system that S82-DG-1 would form part of. The relatively high velocity of S82-DG-1 with respect to NGC 1211 (142 km s$^{-1}$) makes their expected interaction low. The redshift-space ($s$) between both galaxies is $\approx$ 1.9 Mpc. The naive calculation of the crossing time of S82-DG-1 at this distance gives approximately 13 Gyr, so if the galaxies have interacted, only a fly-by is expected. Additionally, quenching mechanisms as tidal stirring \citep{2009Natur.460..605D,2011ApJ...726...98K} are not expected. The tidal radius (more properly the Jacobi radius $r_J$) can be estimated as:
$$ (r_J /r )^3 \; = \; M_{S82-DG-1} / 3 M_{NGC 1211}(<r) \; , $$
where $r$ is the separation between the two galaxies, $M_{S82-DG-1}$ is the total mass of S82-DG-1, and $M_{NGC 1211}(<r)$ is the total mass enclosed within $r\sim 2$ Mpc. Adopting $M_{NGC 1211}(r<2 \mathrm{Mpc}) = 2\times10^{12}$ \Msun, and $M_{S82-DG-1} \sim 10^{10}$ \Msun, the Jacobi radius would be $r_J \sim 240$ kpc, over two orders of magnitude larger than the effective radius. Tidal effects produced by NGC 1211 on this UDG are therefore negligible. Additionally, if a fly-by interaction has occurred, some byproduct is expected, such as the presence of tidal features. We do not find evidences of such features. Nevertheless, such structures are expected to have extremely low surface brightnesses and could be not detectable under the depth limits of our dataset. All these arguments do not allow us to address about the interaction between S82-DG-1 and NGC 1211, so it remains uncertain. Future deeper observations, both in the optical or HI, could shed some light about the issue.

The use of globular cluster systems to obtain distances to extremely low surface brightness galaxies can be extremely useful in the discovery of new objects in low density environments. However, it is necessary to investigate in greater depth what are the properties of the globular cluster systems of these galaxies. Depending on the universality of the GCLFs found, this could support this method as effective. In this sense, it is important to identify diffuse objects in low density environments where their GCLF can be studied with accuracy without the presence of interloping GCs in projection that may be associated to the cluster environment or other nearby massive galaxies. A perfect sample would be those detected by \cite{2018ApJ...857..104G}, in which 781 low surface brightness galaxies are found in a wide area ($\sim$ 200 deg$^2$) using the Hyper Suprime-Cam Subaru Strategic Program (HSC-SSP). Given the great depth of the survey for point sources, and the large angular sizes of the detected objects (effective radius between 2.5 and 14 arcseconds), it is expected the detection of the GCLF for a large fraction of nearby objects. In fact, in a spectroscopic follow-up carried out by \cite{2018ApJ...866..112G}, distances of 25 Mpc and 41 Mpc are obtained for two blue objects which are, therefore, amenable to the detection of GCs. The systematic study of the GCLF for the galaxies of this sample would be an excellent starting point for the accurate characterization of the GC systems properties of diffuse galaxies. It could establish an excellent statistical  catalog for the huge number of extremely faint objects, inaccessible to spectroscopy, that are expected to be detected with the imminent arrival of the LSST.

It is also worth briefly discussing the exploitation of this technique for different surveys. In this work, we have used data from the IAC Stripe82 Legacy Survey \citep{2016MNRAS.456.1359F}. However, this dataset is not particularly competitive in the detection of point sources. This imposes a limitation of the method to distances of up to approximately 30 Mpc, as it is the case of this work. Nevertheless, the potential of this method is expected to be greater when applied to deeper and higher resolution datasets. For instance, surveys such as the Hyper Suprime-Cam Subaru Strategic Program \citep[HSC-SSP; ][]{2018PASJ...70S...4A} or the under development Large Synoptic Survey Telescope \citep[LSST; ][]{2009arXiv0912.0201L} are designed to reach a 5$\sigma$ point source detection of $\sim$27.5 mag in the \textit{r} band, $\sim$3 mag deeper in the detection of point sources than the data presented in this work. This would imply the detection of GCs up to a distance modulus of (m-M)$\approx$ 35 mag, equivalent to 100 Mpc. However, the seeing-limited resolution at such distances could be a serious drawback. Assuming a median seeing of 0.6", the resolution at 100 Mpc would be half of that presented in this work. It is therefore unknown what are the achievable limits for the detection of GCs for the particular conditions of each dataset. Such a study is beyond the scope of our work. Nevertheless, for objects with similar distances to that presented here, considerably higher signal to noise, better resolution and  multi-band detection including \textit{u} and/or infrared bands, would result in a much higher accuracy in the obtained distances. This would both reduce the number of interlopers and improve the photometric uncertainties, greatly increasing the potential of this technique.

\section*{Acknowledgments}

This work has been possible using observations from the GTC telescope, in the Spanish Observatorio del Roque de los Muchachos of the Instituto de Astrof\'isica de Canarias, under Director's Discretionary Time. {We thank the anonymous referee for a careful review and useful comments.} We thank Ignacio Trujillo for crucial discussions that helped the development of this work. We thank Antonio Cabrera for his help and support with the observations and Alexandre Vazdekis for a visual identification of the main features of the spectrum. We thank Mireia Montes, Lourdes Verdes-Montenegro and Michael Jones for helpful discussions. We also thank the participants of the Workshop "The Bewildering Nature of Ultra-diffuse Galaxies" hosted in the Lorentz Center, Leiden, for fruitful and enjoyable conversations. The authors of this paper acknowledges support from grants AYA2014-56795-P and AYA2016-77237-C3-1-P from the Spanish Ministry of Economy and Competitiveness (MINECO). JR thanks MINECO for financing his PhD through FPI grant. M.A.B. acknowledges financial support from the Severo Ochoa Excellence programme (SEV-2015-0548). This work has made use of \texttt{python} \url{(www.python.org)} and \texttt{IRAF} \url{(http://iraf.noao.edu/)} software.


\appendix

\section{Effects of incompleteness in the luminosity function}\label{sec:App_sim}

\begin{figure*}
  \centering
   \includegraphics[width=\textwidth]{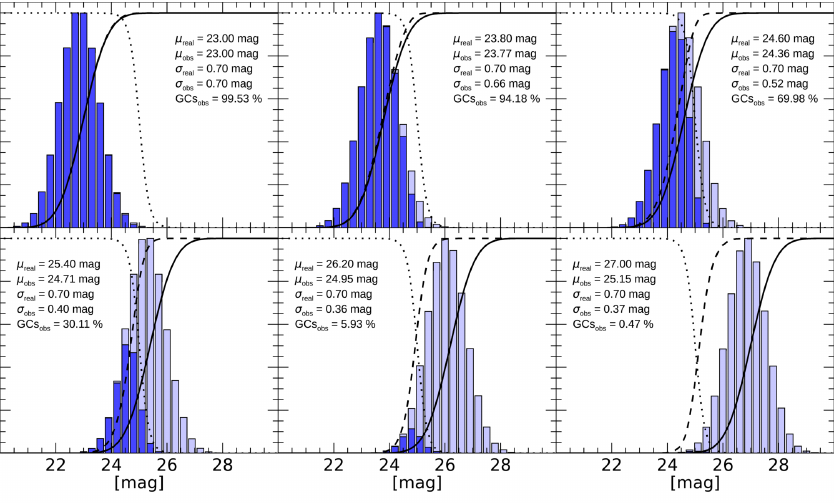}
    \caption{Illustrative scheme of the effect of completeness on the luminosity function of globular clusters. The real luminosity function is plotted in a light blue histogram while the observed one is plotted in dark blue. The line of completeness is shown with a dotted line and and corresponds with the values obtained for the S82-DG-1 field in \textit{g} the band. The dashed line curve represents the cumulative histogram of the observed distribution while the continuous line represents the cumulative histogram of the real distribution. The text shows the properties of the real and observed luminosity functions.}

   \label{fig:LF_test}
\end{figure*}

In this appendix we recreate a simulation to visualize the effects of incompleteness in a globular clusters luminosity function and the recovered observational parameters under different incompleteness regimes. For this we create a Gaussian distribution with a width of 0.7 mag and an arbitrarily high number of GCs to avoid shot noise effects. Next, we apply the effects of incompleteness to this distribution. For this we assign a probability to each GC of not being observed, which will depend on its magnitude and the parameters of completeness. We recreate this effect for the \textit{g} band. In Fig. \ref{fig:LF_test} we plot the real distribution (light blue) and the observed distribution (dark blue) for different peaks of the luminosity function in a range between 23 mag (totally observed luminosity function) and 27 mag (almost non-observed luminosity function). We also plot the cumulative histograms, with an arbitrarily small cumulative bin, for the case of the real luminosity function (solid line) and the observed luminosity function (dashed line). The completeness curve is also plotted with a dotted line. We perform a Gaussian function fitting to the observed luminosity function for the different cases. In each panel, the fitted peak, width and number of globular clusters of the observed distribution are written. As expected, there is a bias in which the peak of the observed luminosity function is systematically brighter than the real one. Additionally the width of the observed distribution is systematically narrower than the real distribution. 


\end{document}